# Evidence for In-Plane Tetragonal c-axis in $Mn_xGa_{1-x}$ Thin Films using Transmission Electron Microscopy

J. Karel[1*], F. Casoli[2‡], P. Lupo[2], L. Nasi[2], S. Fabbrici[2,3], L. Righi[2,4], F. Albertini[2], C. Felser[1]

[1]Max-Planck-Institut für Chemische Physik fester Stoffe, Dresden, Germany 01187
[2]IMEM-CNR, Parma, Italy 43124
[3]MIST E-R Laboratory, Bologna, Italy 40129
[4]Dipartimento di Chimica, Università di Parma, Parma, Italy 43124

[*]julie.karel@cpfs.mpg.de
[‡]francesca.casoli@imem.cnr.it

**Abstract**

Tetragonal $Mn_xGa_{1-x}$ ($x$=0.70, 0.75) thin films grown on $SrTiO_3$ substrates at different temperatures and thicknesses exhibit perpendicular magnetic anisotropy with coercive fields between 1-2T. Transmission electron microscopy (TEM) and X-ray diffraction (XRD) reveal that 40nm samples grown at 300-350$^o$C lead to polycrystalline films with the tetragonal c-axis oriented primarily perpendicular to the film plane but with some fraction of the sample exhibiting the c-axis in the film plane. This structure results in a secondary magnetic component in the out of plane magnetization. Growth at 300$^o$C with a reduced thickness or Mn concentration significantly decreases the presence of the tetragonal c-axis in the film plane, thus improving the magnetic properties. TEM is of critical importance in characterizing these materials, since conventional XRD cannot always identify the presence of additional crystallographic orientations although they can still affect the magnetic properties. Our study points to ways that the microstructure of these thin films can be controlled, which is critical for utilization of this material in spintronic devices.

Mn$_y$Ga ($y$=2,3) in the metastable tetragonal D0$_{22}$ phase exhibits unique properties, which can be exploited for numerous spintronic devices such as spin-transfer-torque RAM.[1] This phase is also interesting for the development of permanent magnets with maximum energy products halfway between Nd$_2$Fe$_{14}$B and Ba(Sr)Fe$_{12}$O$_{19}$.[2] The low saturation magnetization value is due to the ferrimagnetic alignment of Mn on two inequivalent sublattices in the structure and can be tuned by changing the Mn content (i.e.100 kA/m for Mn$_3$Ga to 400 kA/m for Mn$_2$Ga).[3,4,5,6,7] The material also exhibits high uniaxial magnetic anisotropy, a high magnetic Curie temperature and spin polarization.[3,4,5,6,8,9] Mn$_3$Ga crystallizes in three different structures. The hexagonal D0$_{19}$ phase (antiferromagnetic) is the equilibrium phase, while the tetragonal D0$_{22}$ (ferrimagnetic) and the cubic C1$_b$ (ferrimagnetic) phases are mestastable; the latter has only been realized with thin film growth.[10,11,12] This report focuses on the tetragonal D0$_{22}$ phase, which is comprised of two Mn sublattices occupying the 2b and 4d Wyckoff positions. Ga is present on the 2a Wyckoff position. Mn atoms in the 2b position couple ferromagnetically with a moment of 3$\mu_B$ and antiferromagnetically to Mn in the 4d sites (2$\mu_B$), leading to an overall ferrimagnet.[3] Mn on the 4d positions exhibits strong uniaxial anisotropy, with the magnetization pointing along the tetragonal c-axis. Neutron diffraction measurements on bulk powders revealed Mn on the 2b site exhibits a non-collinear ground state with a magnetic easy plane perpendicular to the tetragonal c-axis.[3]

Due to the recent interest in spintronics applications and the metastability of the tetragonal phase, some thin film deposition investigations have been reported. Mn$_3$Ga can be grown epitaxially on MgO and SrTiO$_3$ substrates; Cr, Pt, Ru and Mo buffer layers on MgO have also been successful in obtaining epitaxy.[3,6,13] In addition to reducing the lattice mismatch, buffer layers can also prevent film oxidation, as was recently shown in Mn$_3$Ga thin films with Mg buffer layers on MgO.[14] Epitaxial film growth leads to strong uniaxial magnetic anisotropy with the easy axis out of the plane of the film and coercivity values between 1-2 Tesla.[3,6,13] Some of these reports also find a kink in the out-of-plane hysteresis loop close to *H=0*.[6,9,15] This secondary magnetic component has been attributed either to chemical disorder,[15] to interfacial canting or to alloying with seed layers resulting in superparamagnetic islands in the initial growth phase.[6,13] In this work, a similar out of plane secondary component is also observed and attributed to regions where the tetragonal c-axis is oriented in the plane of the film based on high-resolution transmission electron microscopy results.

Mn$_x$Ga$_{1-x}$ thin films ($x$=0.70 and 0.75) were grown by RF sputtering from MnGa and Mn targets powered at 700 V and using an Ar pressure between 1.5x10$^{-2}$ and 8x10$^{-3}$ mbar; the individual sputter rates were adjusted to obtain the desired composition. The films (20-40 nm) were grown on SrTiO$_3$ substrates at 300 – 350 $^o$C with a 5nm Pt capping layer deposited at room temperature to prevent oxidation. Four samples were investigated: (i) $x$=0.70, 300$^o$C, 40nm, (ii) $x$=0.75, 350$^o$C, 40nm, (iii) $x$=0.75, 300$^o$C, 40nm and (iv) $x$=0.75, 300$^o$C, 20nm. Characterization was performed using X-ray diffraction (XRD), high resolution transmission electron microscopy (HRTEM), SQUID magnetometry and atomic force microscopy (AFM).

Figure 1a shows the typical $\theta$-$2\theta$ XRD pattern for the $Mn_xGa_{1-x}$ films (x=0.70, 0.75) with different growth temperatures, compositions and thicknesses, as indicated; all samples display the $D0_{22}$ structure with the c-axis predominately oriented out of the plane of the film. Figure 1b shows an azimuthal ($\phi$) scan around the off-axis 112 peak; these data exhibit the expected four-fold symmetry, indicating the absence of mosaicity in the films. Figure 1c is a high resolution $\theta$-$2\theta$ XRD scan around the 200 $SrTiO_3$ substrate peak. The 40 nm samples grown at 300 and 350°C with x=0.75 display a shoulder (as indicated) arising from the 200 tetragonal $Mn_3Ga$ phase, although less pronounced in the 300°C sample. This result reflects a portion of the sample where the tetragonal c-axis is oriented in the film plane. Reducing the thickness (to 20nm) or the Mn concentration evidently improves the film quality since no such shoulder is observed. The out of plane lattice constant (c) slightly decreases with increasing x (7.14 Å x=0.70 to 7.12 Å x=0.75), whereas the in-plane lattice constant (a) is around 3.90Å for all samples, with no clear x dependence. Consistent with these results, a recent theoretical work predicted a decrease in c with increasing x and no strong dependence on composition for a,[16] and indeed experimental reports reflect this trend.[3,4,7]

Figure 2 shows magnetization (M) perpendicular to the film plane (referred to as out of plane) and in-plane versus applied magnetic field (H) for all samples. The overall magnetic moment decreases when x increases, consistent with previous reports[3,4,5,6] and arises from the antiferromagnetic alignment of Mn between the Wyckoff 2b and 4d positions. Moreover, a large coercivity (between 1-2 T) is observed in the out of plane direction. The $K_1$ values for the x=0.75, 300°C and x=0.70, 300°C samples are consistent with previous reports[3] and roughly estimated to be in the range 7 – 9 x$10^5$ MJ/$m^3$. The maximum energy product $(BH)_{max}$ has been evaluated for the x=2.6, 300 °C sample, i.e., for the composition showing the highest saturation magnetization and therefore most interesting for realizing new permanent magnets. The $(BH)_{max}$ value is 8.7 kJ/$m^3$, slightly less than the values reported in reference [4] for bulk samples. A secondary magnetic component in the out of plane M(H) curve and an in-plane hysteresis are observed most prominently for the x=0.75, 350°C and 300°C samples (subfigures a,b) and reduced with decreasing thickness or Mn concentration (subfigures c,d). A small amount of a secondary magnetic phase is also evident in the out of plane and in-plane M(H) curves of the x=0.70, 300°C sample; similar results were observed in previous reports.[6,8,13] Such features in the M(H) curves are also consistent with a different orientation of the tetragonal phase. High resolution TEM was performed to further investigate this point.

Figure 3a shows a cross sectional HRTEM image from the x=0.75, 350°C sample with the Fast Fourier Transform (FFT) of the film in the inset. Diffraction spots for tetragonal $Mn_3Ga$ with the two different orientations (200) and (004), parallel and perpendicular to the film plane respectively, are highlighted. The color map in Figure 3b displays the spatial distribution of the two different crystallographic orientations giving rise to the diffraction spots shown in the FFT. These results show that the tetragonal c-axis is oriented in the plane of the film for a significant portion of the sample, as shown in yellow in Figure 3b, although the lattice mismatch ($Mn_3Ga$=3.57 Å (c/2) versus $SrTiO_3$=3.90 Å) is larger in this orientation. Moreover, evidence for the hexagonal phase in small amounts near the film surface was also found (not shown here). These data indicate the sample is polycrystalline and not single phase, with grains of irregular shape and size, as shown in the inset of Figure 3b. By contrast, Figure 3c shows the HRTEM image of a single grain of $Mn_3Ga$ grown at 300°C. The FFT of the whole area (Figure 3d) exhibits reflections from a single crystallographic orientation of the film corresponding to the tetragonal $Mn_3Ga$ with c-axis oriented perpendicular to the film plane, in addition to the substrate reflections. The inset of figure 3d further shows the film forms faceted islands with a single out of plane orientation. We found that a very small portion of the sample (less than 2%) exhibited the c-axis in the film plane (not shown).

Figure 4 displays the AFM images from the (a) $x$=0.75, 350°C and (b) $x$=0.70, 300°C samples; the RMS roughness is 7.1 and 5.5nm, respectively. The island-like morphology of the samples accounts for these high RMS roughness values, while on the island surface the roughness is reduced. The better crystal structural quality at lower temperature allows the formation of larger islands (Fig. 4(b)). Additionally, the islands show a well-defined shape with facets oriented along specific crystallographic directions. Upon comparison, it is clear that the sample grown at elevated temperature is generally poorer in quality than the film grown at reduced temperatures, in agreement with the XRD and TEM results. The image from the $x$=0.70, 300°C sample shows a film that is smoother and more uniform, consistent with a primarily single phase epitaxial film.

The $M(H)$ results can be interpreted coherently with the XRD and TEM measurements from this work. The in-plane hysteresis of the $x$=0.75, 350 °C sample is due to the sample portion with the c-axis oriented in the film plane and is reduced with decreasing growth temperature and thickness (Fig. 2a,b,c). The kink evident in the out-of-plane loop close to $H=0$ shows a similar trend. Both of these features are further reduced in the sample $x$=0.70, 300 °C (Fig. 2d); however it does not show a perfectly squared perpendicular loop. This loop is instead typical of a composite sample, with a small fraction of easy-plane regions, which are coupled through exchange-interaction to the predominant fraction that exhibits a perpendicular easy-magnetization direction.[17] Also the magnetization values at 5 T ($M_{5T}$) for the four samples analyzed in this work are consistent with the structural investigation and with the two magnetic sublattice model previously proposed for the $D0_{22}$ structure.[3,4,16,18] As stated above, $M_{5T}$ increases with decreasing Mn concentration, from 125 kA/m at $x$=0.75 to 225 kA/m at $x$=0.70. A preferential removal of Mn from the 2b sites with decreasing $x$ can be deduced from these magnetization values. In fact, atomic moments for the two sublattices consistent with previously published neutron data and calculations,[3,4,18] i.e., ≈3$\mu_B$/atom for 2b sites and ≈2$\mu_B$/atom for 4d sites, can only be obtained from the above experimental $M_{5T}$ by taking into account a preferential depletion of the 2b sites and a nearly full occupancy of the 4d sites in the $x$=0.70 sample. This is consistent with the occupancy models employed in references 3 and 16. The lower $M_{5T}$ measured on the sample grown at 350°C compared to the samples grown at 300°C with the same composition reflects the larger fraction of in-plane c-axes and occurrence of the antiferromagnetic $D0_{19}$ phase at 350°C.

In summary, $Mn_xGa_{1-x}$ thin films ($x$=0.70, 0.75) were grown at different temperatures and thicknesses on $SrTiO_3$ substrates. Samples grown at 300 - 350°C evidently lead to polycrystalline films with a significant fraction exhibiting the tetragonal c-axis in the film plane, which subsequently results in a secondary magnetic component in the out of plane $M(H)$ curve as well as an in-plane hysteresis. Reducing the film thickness or Mn concentration improves the film quality; the 300°C $x=0.75$ 20nm sample shows only a small amount of the c-axis in plane, whereas the $x=0.70$ sample exhibits primarily a single crystallographic orientation. Perpendicular magnetic anisotropy and thus potential utilization of this material in future spintronic applications depends on carefully controlling the film structure with the appropriate growth conditions. Other reports have also evidenced the secondary magnetic component in the out of plane hysteresis loops and attributed it to other origins. We showed here it can prove challenging to identify the presence of this additional crystallographic orientation using conventional XRD although it can still affect the magnetic properties. Therefore, TEM is of critical importance in characterizing these materials.

**Figure Captions**

**Figure 1. (a)** $\theta$-$2\theta$ XRD pattern $Mn_xGa$ films ($x$=0.70, 0.75) with different growth temperatures as indicated. All thicknesses are 40 nm, unless otherwise noted. The grey squares (purple squares) identify peaks from the substrate (Pt cap). **(b)** Azimuthal ($\phi$) scan around the off-axis 112 peak. **(c)** High resolution $\theta$-$2\theta$ XRD near the 200 $SrTiO_3$ substrate peak with the $Mn_3Ga$ 200 peak indicated. Colors are the same for all subfigures.

**Figure 2.** In-plane (solid) and out-of-plane (dashed) magnetization at room temperature versus applied magnetic field for **(a)** $x$=0.75, 350°C, **(b)** $x$=0.75, 300°C, **(c)** $x$=0.75, 300°C 20 nm and **(d)** $x$=0.70, 300°C. Samples are 40 nm thick, unless otherwise noted.

**Figure 3. (a)** Cross sectional HRTEM image from the $x$=0.75, 350°C sample. The inset shows a portion of the FFT in the film region. Diffraction spots for tetragonal $Mn_3Ga$ with 2 different orientations (200 and 004) are highlighted. **(b)** Color map showing the spatial distribution of the different crystallographic orientations highlighted in the FFT. The inset shows the film microstructure **(c)** HRTEM image and **(d)** corresponding FFT from the $x$=0.75, 300°C sample. The inset shows that the film forms faceted islands with a single out of plane orientation.

**Figure 4.** AFM images of **(a)** $x$=0.75, 350°C and **(b)** $x$=0.70, 300°C samples. Each image is approximately 1μm x 1μm; the color scale corresponding to the height is different for each image.

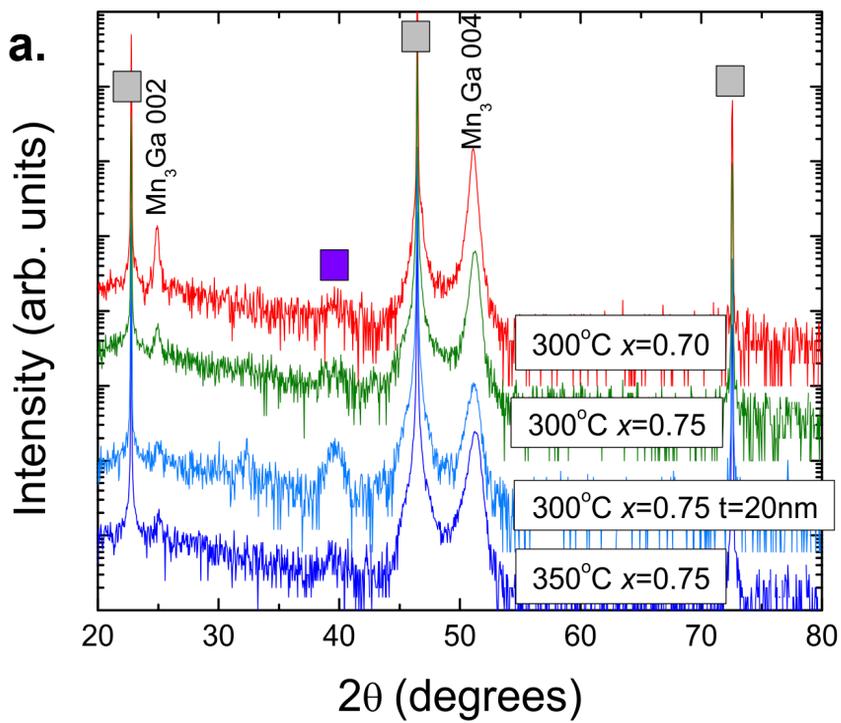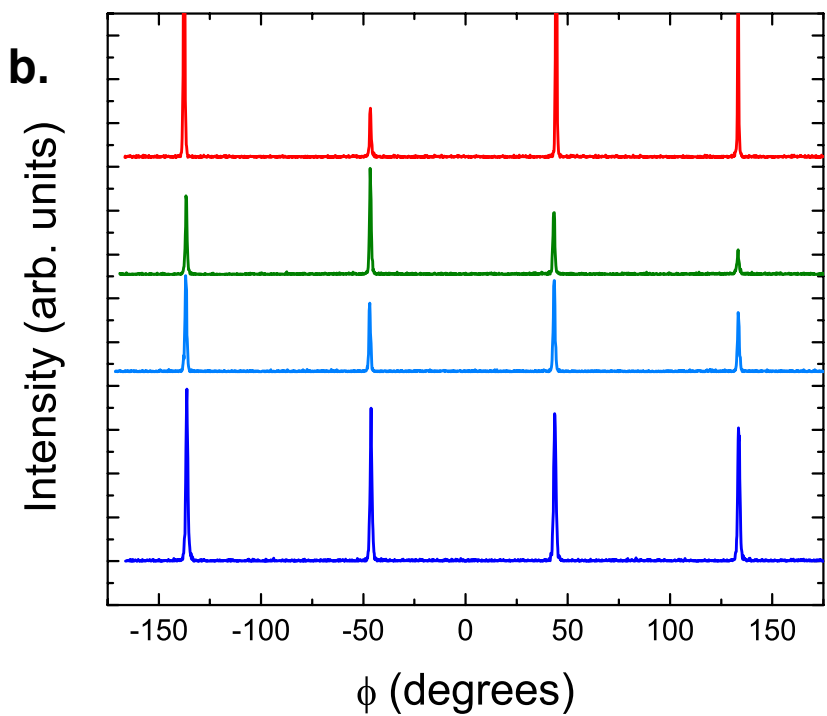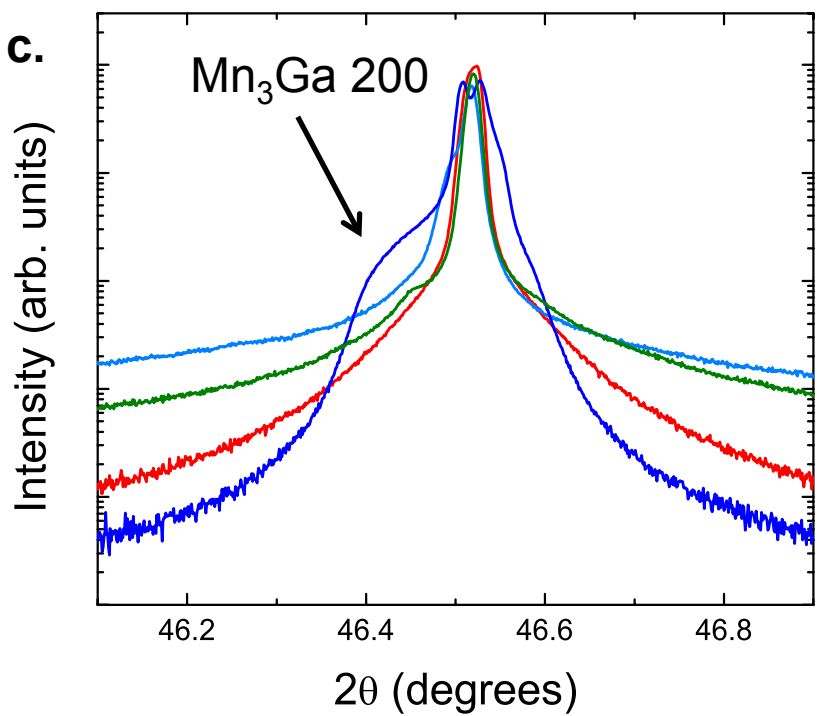

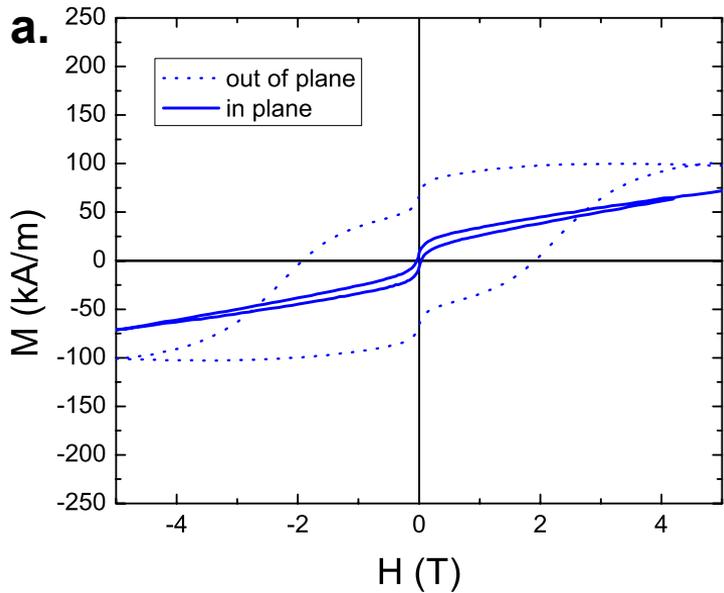 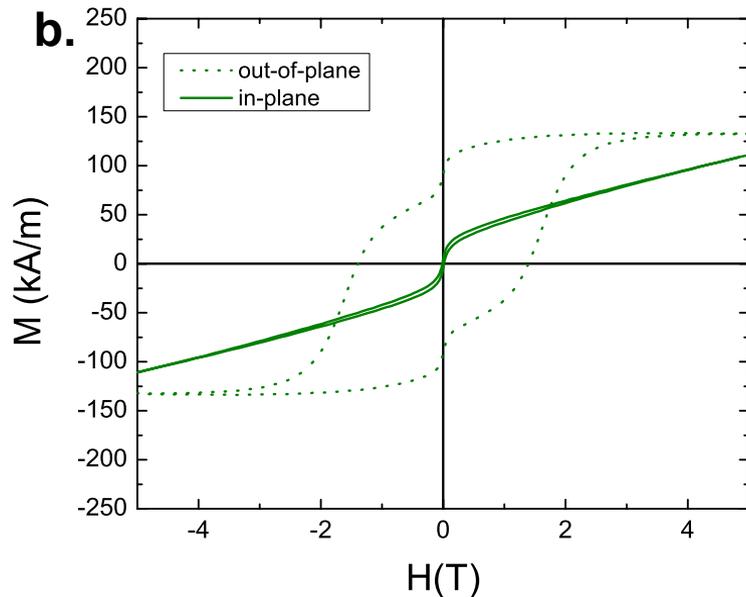
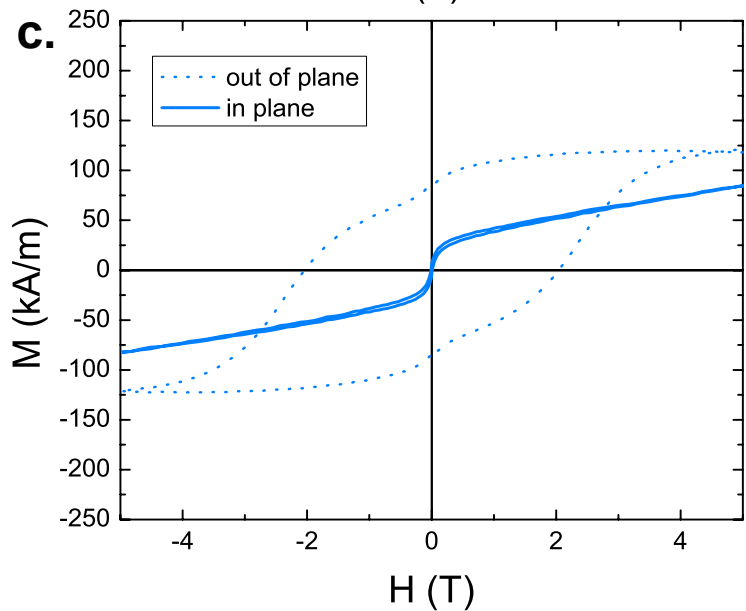 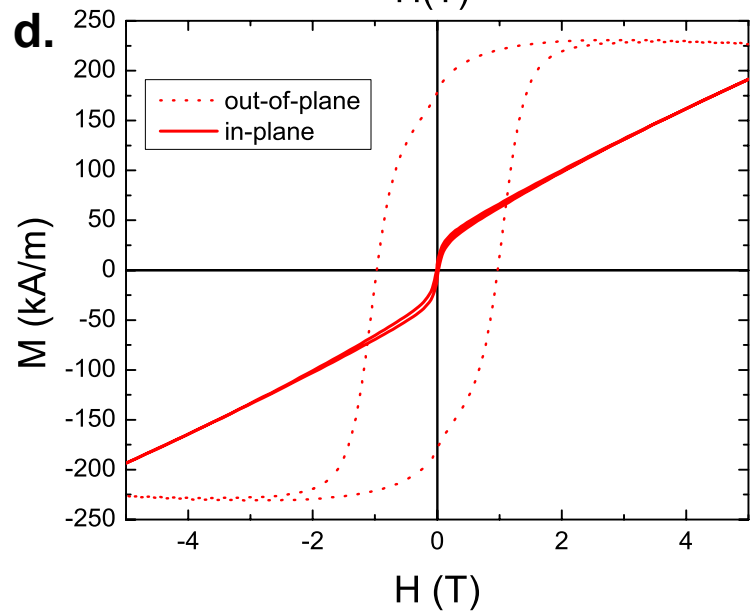

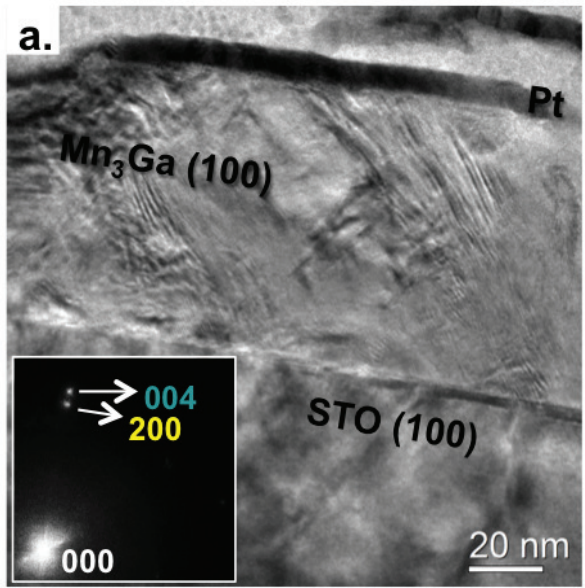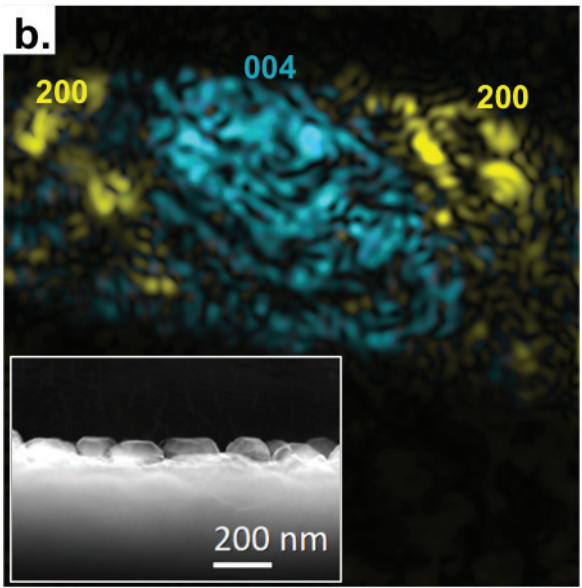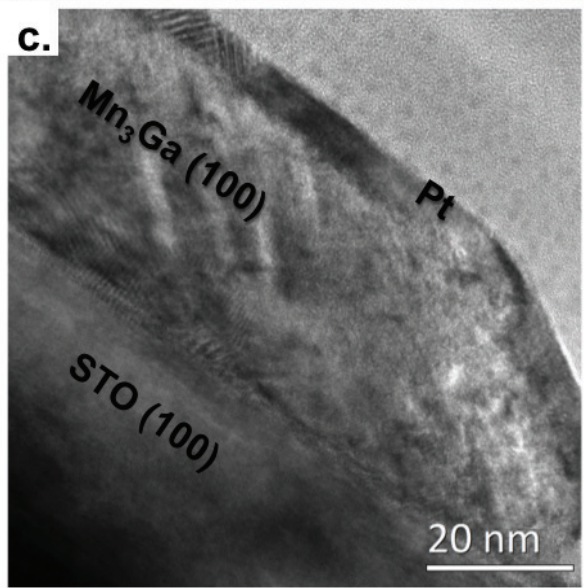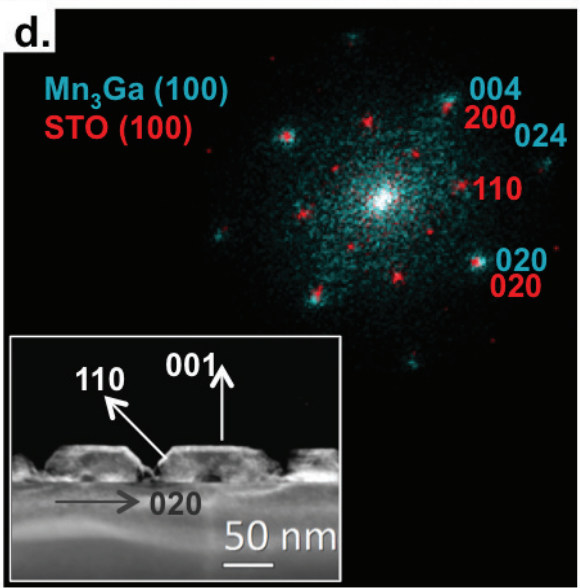

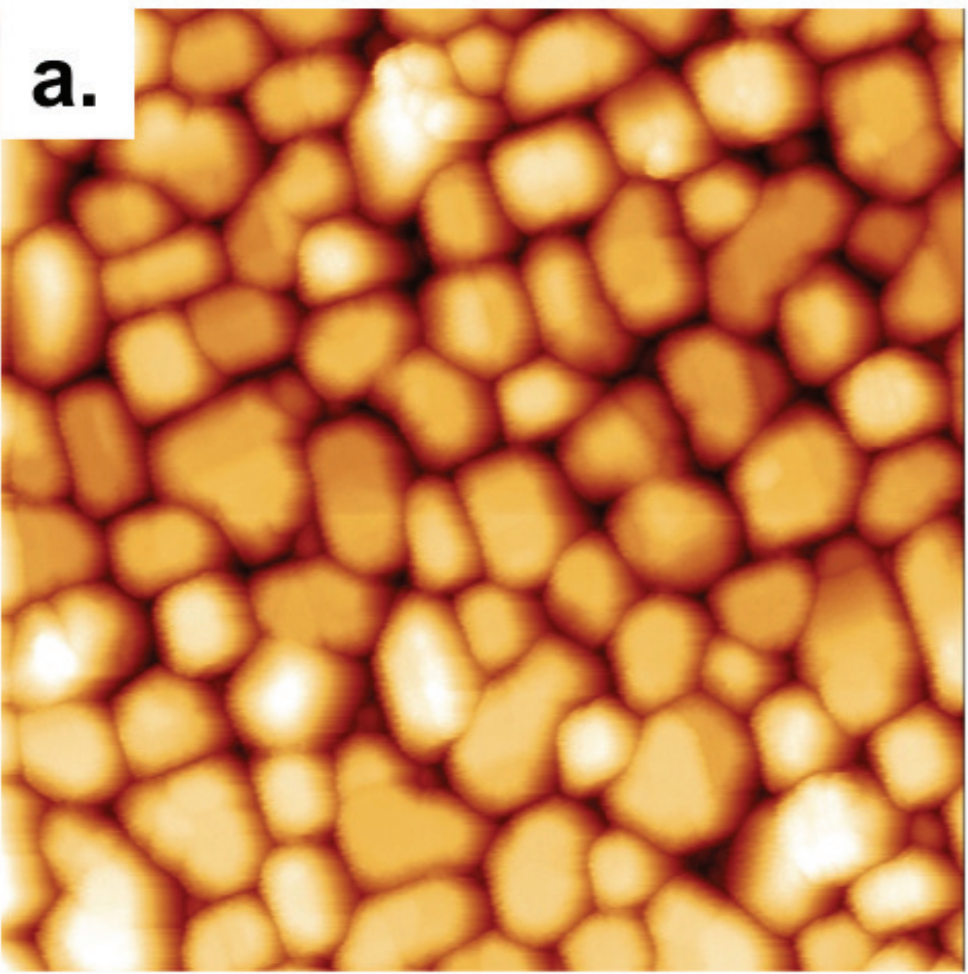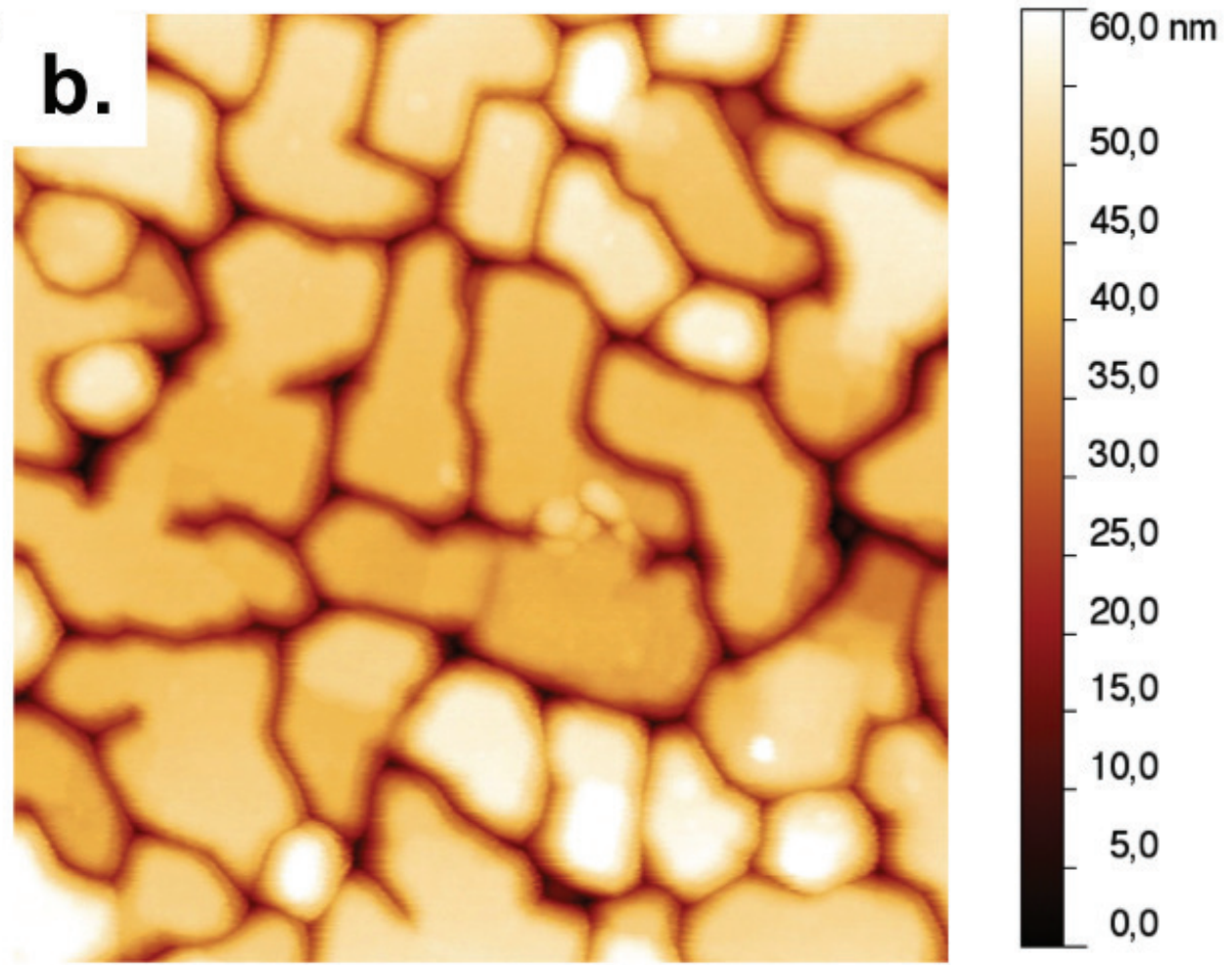